# Comparison of three reconstruction algorithms for low-dose phase-contrast computed tomography of the breast with synchrotron radiation

S. Donato[1,2,3], S. Caputo[4], L. Brombal[5,6], B. Golosio[7,8], R. Longo[5,6], G. Tromba[9], R.G. Agostino[2], G. Greco[1], B. Arhatari[10], C. Hall[10], A. Maksimenko[10], D. Hausermann[10], D. Lockie[11], J. Fox[12], B. Kumar[12], S. Lewis[13], P.C. Brennan[13], H.M. Quiney[14], S.T. Taba[13] and T.E. Gureyev[14]

[1] Department of Mathematics and Computer Science, University of Calabria, 87036 Rende (CS), Italy
[2] Department of Physics and STAR-Lab, University of Calabria, 87036 Rende (CS), Italy
[3] INFN-LNF, 00044 Frascati (RM), Italy
[4] Department of Physics, University of Milano-Bicocca, Milano, Italy
[5] Department of Physics, University of Trieste, I-34127 Trieste (TS), Italy
[6] INFN Division of Trieste, I-34127 Trieste (TS), Italy
[7] Department of Physics, University of Cagliari, I-09042 Monserrato (CA), Italy
[8] INFN Division of Cagliari, I-09042 Monserrato (CA), Italy
[9] Elettra-Sincrotrone Trieste S.C.p.A., I-34149 Trieste (TS), Italy
[10] Australian Synchrotron, ANSTO, Clayton 3168, Australia
[11] Maroondah BreastScreen, Ringwood East 3135, Australia
[12] Monash University, Clayton 3800, Australia
[13] The University of Sydney, Lidcombe 2141, Australia
[14] The University of Melbourne, Parkville, 3010, Australia

**Abstract**

Three different computed tomography (CT) reconstruction algorithms: Filtered Back Projection (FBP), Unified Tomographic Reconstruction (UTR) and customized Simultaneous Algebraic Reconstruction Technique (cSART), have been systematically compared and evaluated using experimental data from CT scans of ten fresh mastectomy samples collected at the Imaging and Medical beamline of the Australian Synchrotron. All the scans were collected at the mean glandular dose of 2 mGy, using monochromatic X-rays with 32 keV energy, flat-panel detectors with 0.1 mm pixels and 6 meter distance between the rotation stage and the detector. Paganin's phase retrieval method was used in conjunction with all three CT reconstruction algorithms. The reconstructed images were compared in terms of the objective image quality characteristics, including spatial resolution, contrast, signal-to-noise and contrast-to-noise ratios. The images were also evaluated by seven experienced medical imaging specialists, rating perceptible contrast, sharpness of tissue interfaces, image noise, calcification visibility and overall image quality. Of the three compared algorithms, cSART was clearly superior to UTR and FBP in terms of most measured objective image quality characteristics. At the same time, the results of the subjective quality evaluation consistently favoured the images reconstructed by FBP, followed by UTR, with cSART receiving lower scores on average. We argue that this apparent disagreement between the objective and subjective assessments of image quality can be explained by the importance assigned to image contrast in the subjective assessment, while the signal-to-noise ratio seemed to receive relatively low weighting. This study was conducted in preparation for phase-contrast breast CT imaging of live patients at Australian Synchrotron (Melbourne, Australia).

## 1. Introduction

Breast cancer is a complex and significant health issue affecting millions of people worldwide (Sung et al., 2021). It is the most common cancer among women globally and can also occur in men, although much less frequently (Fox et al., 2022). Breast cancer, like many forms of cancer, arises when cells in the breast tissue begin to grow uncontrollably (Sledge and Miller, 2003). These cells can form a tumor, which may be benign (non-cancerous) or malignant (cancerous). Detecting breast cancer early is crucial for improving outcomes. Regular screening mammograms can help detect abnormalities in the breast tissue before symptoms develop (Autier and Boniol, 2018). While the implementation of these screening programs has successfully demonstrated a decrease in mortality rates, there is significant potential for further improvement in the diagnostic accuracy of breast imaging. If a suspicious finding is detected, further diagnostic tests, such as ultrasound, MRI, or biopsy, may be performed to confirm the presence of cancer (Jafari et al., 2018; Vaughan, 2019). Two-dimensional (2D) mammography has inherent limitations, including low soft tissue contrast and the overlap of different tissues in 2D X-ray projections, which can hinder detecting masses, suspicious lesions or cysts. Overcoming the overlap of diagnostically relevant tissue structures remains a critical challenge in producing clinically valuable breast images. Additionally, there is a potential radiation risk associated with screening mammography (Feig and Hendrick, 1997), along with significant discomfort for patients due to breast compression (Boone et al., 2001). Digital breast tomosynthesis (DBT) (Chong et al., 2019) and dedicated breast computed tomography (bCT) (Sarno et al., 2015) are newer imaging technologies aimed at addressing the superimposition challenges in 2D mammography. Current data indicate both advantages and disadvantages of DBT and bCT compared to 2D mammography, with no significant dose reduction observed for either technique and limited reduction in discomfort for DBT.

The majority of currently available X-ray-based breast imaging methods primarily rely on X-ray attenuation data, which is based on the differing absorption properties of various soft tissues. However, this approach does not offer substantial contrast for soft tissues due to the minor density variations among different components of breast tissue. Phase-contrast (PhC) imaging represents a more advanced X-ray technique capable of capturing X-ray wave refraction and phase shifts as they pass through objects (Momose et al., 1996; Taba et al., 2018). Particularly with high-energy X-ray beams, phase shifts can provide significantly stronger contrast than attenuation alone. Consequently, retrieving the phase shift of X-ray beams holds great potential for enhancing breast image quality. Various PhC imaging techniques exist, such as propagation-based imaging (PBI) (Gureyev et al., 2019), analyzer-based imaging (Keyriläinen et al., 2011), crystal interferometry (Ingal & Beliaevskaya 1995; Davis et al., 1995), edge illumination (Olivo, 2021), and grating interferometry (Hellerhoff et al., 2019). Compared to all other phase-contrast techniques, PBI is experimentally the simplest way to exploit phase shift information because it does not require any X-ray optical elements between the sample and the detector. X-ray PBI CT (PB-CT), which utilizes refraction as well as absorption of X-rays in tissue, shows particular promise due to its superior sensitivity to soft tissues, including tumors. However, PBI requires high spatial coherence of the incident X-ray beam in order to render the phase contrast detectable. In this context, synchrotron facilities have been pivotal in advancing PBI, especially in breast cancer research. Synchrotrons offer unparalleled X-ray brilliance and coherence, enabling high-resolution imaging with exquisite detail. This high intensity coherent photon flux enhances the detection of subtle tissue density variations, revealing previously elusive levels of detail (Baran et al., 2018; Gureyev et al., 2019; Longo et al., 2019; Arana Peña et al., 2023). These facilities have fostered collaboration among physicists, pathologists, radiologists, and oncologists, driving breast cancer diagnostics forward (Castelli et al., 2011; Zhao et al., 2012; Gureyev et al., 2019; Longo et al., 2019; Pacilè et al., 2019; Donato et al., 2024a; Donato et al., 2024b).

Ongoing efforts at synchrotrons like Elettra in Italy and Australian Synchrotron focus on refining imaging setups and protocols for potential clinical implementation of PB-CT, aiming for a mean glandular dose (MGD) comparable to clinical mammography or lower (Baran et al., 2017; Longo et al., 2019; Tavakoli Taba et al., 2019; Oliva et al., 2020). The requirement of low MGD can be satisfied by either reducing the X-ray fluence per tomographic projection (Greffier et al., 2015) or by decreasing the number of projections (Sidky et al., 2014). The former method, while maintaining good angular sampling, results in increased noise in the projection images, leading to a noisier CT image. Conversely, reducing the number of projections significantly below the Nyquist angular sampling criterion introduces significant image artifacts and increased noise when using analytical reconstruction algorithms. Various approaches have been proposed to enhance overall image quality in low-dose CT scans, some of which have been applied to breast CT data, including iterative reconstruction (IR) algorithms (Makeev and Glick, 2013; Bian et al., 2014; Pacilè et al., 2015). Within this framework, optimizing the reconstruction algorithm stands as one of the final stages in clinically implementing PB-CT. The goal is to enhance overall image quality of low-dose CT scans suitable for clinical use, aiming for a total MGD below 5mGy.

This study aims to compare two novel CT reconstruction algorithms, the Unified Tomographic Reconstruction (UTR) (Gureyev et al., 2022) and a customized Simultaneous Algebraic Reconstruction algorithm (cSART) (Donato et al., 2022), with the well-established Filtered Back Projection (FBP) algorithm (Natterer, 2001), in the context of breast PB-CT. The goal is to evaluate the performance of these algorithms in terms of objective metrics such as contrast, signal-to-noise ratio (SNR), spatial resolution, and the ratio of SNR to spatial resolution, alongside an assessment of their clinical and radiological image quality via a human observer study. Accurate reconstruction is emphasized as essential for diagnostic purposes, impacting the quality and interpretability of medical images.

## 2. Materials and Methods
### a. Breast tissue specimens

Approval was granted by the Human Research Ethics Committee (project number: CF15/3138 - 2015001340), and the study utilized 10 fresh mastectomy specimens after having obtained written consent from the patients. Samples were not fixed or preserved, and all were scanned shortly after surgical excision. Although most specimens contained in situ and/or invasive tumors, a few did not exhibit malignancy or contained only benign lesions upon subsequent pathology examinations. Basic information about each mastectomy sample, including weight, size, and histopathological diagnosis, is provided in Table 1 alongside some key scan parameters.

*Table 1. The mastectomy samples description*

| Sample Number | Patient | Breast side | Sample dimension (mm) | Sample weight (g) | Diagnosis | Detector | Number of projections |
|---|---|---|---|---|---|---|---|
| 1 | 54 yrs female | Left | 185 x 190 x 65 (ML x SI x AP). | 835 | No in situ malignancy; Invasive carcinoma, mixed ductal and lobular features | Hamamatsu | 1800 |
| 2 | 40 yrs female | Left | 170 x 140 x 25 (ML x SI x AP). | 295 | High grade DCIS; No invasive tumour | Hamamatsu | 1200 |
| 3 | 56 yrs female | Left | 230 x 200 x 70 (ML x SI x AP). | 1200 | Residual intermediate grade DCIS, three foci at previous surgical site, clear of excision margin; No invasive carcinoma | Hamamatsu | 1200 |
| 4 | 46 yrs female | Right | 220 x 160 x 40 (ML x SI x AP). | 773 | Negative for malignancy | Hamamatsu | 1200 |
| 5 | 53 yrs female | Right | 220 x 190 x 40 (ML x SI x AP). | 832 | High grade DCIS; No invasive tumour | Hamamatsu | 1200 |
| 6 | 54 yrs female | Left | 200 x 180 x 55 (ML x SI x AP). | 906 | Post neoadjuvant chemotherapy of invasive carcinoma, no special type: No residual in-situ or invasive disease | Hamamatsu | 1200 |
| 7 | 39 yrs female | Right | 160 x 140 x 45 (ML x SI x AP). | 368 | Post chemotherapy and radiotherapy: intermediate grade DCIS; No residual invasive tumour | Hamamatsu | 1200 |
| 8 | 44 yrs female | Right | 120 x 120 x 20 (MLx SI x AP). | 305 | Residual invasive lobular carcinoma (post chemo- and radio-therapy) | Xineos | 2400 |
| 9 | 53 yrs female | Left | 140 x 145 x 27 (SI x ML x AP). | 342 | Residual tumour bed, No residual invasive tumour | Xineos | 2400 |
| 10 | 49 yrs male | Left | 128 x 25 x 25 (ML x SI x AP). | 234 | Invasive carcinoma, no special type, Grade 2 | Xineos | 2400 |

ML - medial to lateral; SI - superior to inferior; AP - anterior to posterior; DCIS, ductal carcinoma in situ; yrs - years old.

### b. Experimental setup

All tomographic acquisitions for this study were carried out at the Imaging and Medical Beamline (IMBL) of the Australian synchrotron facility in Melbourne. IMBL utilizes a super-conducting wiggler and bent double crystal monochromator system to generate a parallel monochromatic X-ray beam with a cross-section of up to approximately 500 mm (width) × 30 mm (height) in the energy range of 20-120keV, with an energy resolution of $\Delta E/E \cong 10^{-3}$. Two detectors were employed for the scans: 1) a Teledyne-Dalsa Xineos-3030HR flat panel detector with an active area of 296 × 296 mm$^2$ (2988 × 2988 pixel field of view), a pixel pitch of 99 µm and a frame rate of 40 fps, and 2) a Hamamatsu C10900D CMOS flat panel detector, with an active area of 124.8 x 124.8 mm$^2$ (1248 x 1248 pixel field of view), a pixel size of 100 µm, and a frame rate of 17 fps. The two detectors had very similar performance characteristics in terms of quantum efficiency and spatial resolution. During the scan, samples were placed in a thin-walled plastic cylindrical container measuring 11 cm in diameter. Each scan was conducted at a clinically relevant mean glandular dose of 2 mGy, distributed evenly across either 1200, 1800 or 2400 projections with a uniform angular step of 0.15, 0.1 or 0.075 degrees, respectively, over 180 degrees, as detailed in Table 1. As the mastectomy samples, when placed in the cylindrical container, had the height exceeding that of the incident X-ray beam, the CT scans were performed in several (up to 7) increments ("slabs"), with the height of each slab equal to 2.5 cm and the consecutive slabs overlapping vertically by 1 cm. The overlap was used for subsequent stitching of the slabs in the vertical direction into a single scan with the height of projections exceeding that of the imaged sample. Note that the areas of the overlap (at the top and/or bottom of the slabs) did not receive more incident photons than the "central" areas of the slabs. The extra exposure in the overlap areas was compensated by the lower intensity, due to the roll-off of the incident beam which had an approximately Gaussian profile with the standard deviation of ~1 cm in the vertical direction.

The scans employed quasi-plane monochromatic X-rays with an energy of E=32 keV. The distance between the sample and the detector was 6 m in free space, while the source-to-detector distance was 143 m. Dark-current images (with no beam) and flat-field images (with the beam, but without the imaged sample) were collected immediately before and after each CT scan. An ionization chamber was employed to measure the photon fluence rate and the associated air kerma. MGD was subsequently calculated using Monte Carlo simulations (Nesterets et al., 2015). These simulations utilized a numerical phantom representing the breast, comprising 30% glandular tissue and 70% adipose tissue, surrounded by a 5 mm thick layer of tissue to simulate the skin.

### c. Image reconstruction

Image reconstruction was performed using three different algorithms: FBP, UTR and cSART. Before image reconstruction, all projection images were pre-processed through conventional flat field and dark current corrections. Phase-retrieval was also performed using the well-known Paganin's Homogeneous Transport of Intensity Equation (TIE-Hom) algorithm (Paganin et al., 2002). The TIE-Hom algorithm functions as a low-pass filter controlled by a single parameter, denoted as γ. To ensure accurate phase retrieval, eliminating diffraction fringes at material boundaries, the value of γ should match the ratio δ/β, representing the real decrement to the imaginary part of the relative complex refractive index between the two materials (e.g. glandular and adipose tissue, in the case of breast tissue samples) at a given X-ray energy. Adjusting γ allows for trade-offs between signal-to-noise ratio (SNR) and spatial resolution (Gureyev et al., 2017); increasing γ can improve SNR, albeit at the expense of spatial resolution, while decreasing γ enhances spatial resolution but reduces SNR accordingly. In this study, we employed a "half phase retrieval" approach, setting γ to approximately one-half of the theoretical δ/β value for glandular tissue relative to blood at 32 keV, which is equal to 275. This choice of γ was consistent with our previous studies on optimization of image quality in breast PB-CT (Taba et al., 2019). A Hamming filter was employed for FBP reconstructions.

### i. The Unified Tomographic Reconstruction algorithm

The UTR method for three-dimensional reconstruction of objects from transmission images collected at multiple illumination directions was described in (Gureyev et al., 2022). The key features of the UTR algorithm are as follows.

- The UTR method is applicable to experimental conditions relevant to absorption-based, phase-contrast or diffraction imaging using X-rays, electrons and other forms of penetrating radiation or matter waves.
- It unifies the conventional, phase-contrast and diffraction CT models by intrinsically incorporating both the phase retrieval and the correction for the Ewald sphere curvature (in the cases with a shallow depth of field and significant in-object diffraction).
- The numerical algorithm implementing UTR, as used in this study, is based on three-dimensional gridding, allowing for fast computational implementation, including parallel processing of multiple input projection images. In principle, this algorithm can be used with any scanning geometry involving plane-wave illumination.

The software code, implementing the UTR algorithm, that was used in the present study, is publicly available (Gureyev, 2024). The only non-trivial user-defined parameter of the UTR algorithm relevant to the present study was the "noise-to-signal ratio", which corresponded to the inverse of the SNR in input projections. This parameter was set to 0.05 (corresponding to 5% noise) in all the reconstructions. Increasing the value of this parameter results in stronger low-pass filtering and consequent noise suppression in the reconstructed images, at the expense of spatial resolution.

### ii. The custom SART algorithm

As discussed in (Donato et al., 2022), the main features of the cSART algorithm can be summarized as follows:

- A relaxation factor, denoted as $\eta$, is employed to adjust the iterative corrections. This factor aims to reduce image noise during the reconstruction process. In our approach, $\eta$ gradually increases from zero to a maximum value over the initial angular steps, then decreases linearly with both the number of iterations and angular steps until it reaches zero at the final angular step of the last iteration.
- Projections corresponding to various angles are utilized in a random order scheme.
- Additionally, a bilateral 3D filter is periodically applied to the reconstructed image during the iterative process. This filter replaces each pixel's content with a weighted average considering both the 3D Euclidean distance and the gray-level difference of neighbouring pixels.
- The optimization process involves adjusting four parameters: the number of iterations, spatial width of the filter ($\sigma_{xy,z}$), pixel intensity difference width ($\sigma_v$), and a weighting factor ($w$).

For optimizing the cSART parameters, reconstructions of a 1 mm thick slice of the tissue were generated using various combinations of the algorithm's parameters. This involved adjusting $\sigma_{xy}$ and $\sigma_z$ within the range of 1 to 10 pixels with a step of 1 pixel, $\sigma_v$ within the range of 0.01 to 0.20 with a step of 0.01, and $w$ within the range of 0.04 to 0.20 with a step of 0.02. This process resulted in a total of 1800 reconstructions. The number of iterations remained fixed at 5, consistent with a typical SART reconstruction, while the regularization filter was applied every 100 randomly ordered angular steps. Following the optimization discussed in (Donato et al., 2022), a subset of optimal parameters is selected based on a threshold value for the frequency peak of the 1D noise power spectrum compared to the equivalent FBP one, evaluated in uniform region of interests within the adipose tissue. Reconstructions falling within this threshold (difference less than 15%) are then compared in terms of signal-to-noise ratio and spatial resolution. The subset of parameters that yields the best values for both metrics is then chosen to perform the full volume reconstruction. This optimization process was repeated 3 times, once for each different number of acquired projections. For datasets with 1200 and 1800 projections the optimal subset of parameters was: *w* = 0.04, *σ*$_{xy,z}$ = 10, *σ*$_v$ = 0.20. For datasets with 2400 projections was: *w* = 0.06, *σ*$_{xy,z}$ = 10, *σ*$_v$ = 0.07.

### d. Two types of reconstructed slices

The reconstructed slices produced by the three different algorithms were initially produced in the planes corresponding to the coronal view in mammography. For radiological assessments, we digitally reoriented the images into axial view, representing the craniocaudal view commonly used in breast imaging. While maintaining the original in-plane resolution of 100 µm, we employed 30-pixel binning to create thicker slices, each measuring 3 mm, for assessment purposes. The latter binning was performed as follows. Each "column" consisting of 30 pixel values at a fixed transverse, (x,y), location inside a stack of 30 adjacent original 100 µm slices was sorted into two bins, the "lower" bin containing all pixel values lower or equal to the selected threshold value $\beta_{Thresh} = 2.e-10$ and the "upper" bin containing the pixels values higher than the threshold value. This particular threshold value was chosen on the basis of analysis of many reconstructions of mastectomy samples, collected at IMBL with plane monochromatic X-rays with *E* = 32 keV, as an optimal threshold between the $\beta$ values of soft (adipose and glandular) tissues and calcifications. If the upper bin was empty (indicating the absence of calcification at this location), the values in the lower bin were averaged, producing a single "denoised" soft-tissue value of $\beta$ for the resultant 3-mm thick slice. If the upper bin was not empty (i.e. a calcification was present), the output "denoised" calcification value of $\beta$ was made equal to the average pixel value in the upper bin only. This procedure was developed in collaboration with radiologists and employed here in order to avoid averaging of $\beta$ values of microcalcifications with those of z-adjacent pixels containing soft tissues. The 3 mm axial slices were produced with a 1.5 mm step, i.e. with a 15-pixels overlap of the 30-slice stacks of original thin slices. This overlap was created in order to reduce apparent "jumps" in the appearance of consecutive 3 mm thick slices during the subjective radiological evaluation. For the evaluation, the 3 mm axial slices were saved in DICOM files, together with sufficient information about the sample, to enable convenient examination of these files on medical PACS systems. For the DICOM files, the data was also converted from the original 32-bit floating-point to a 12-bit integer format by a linear mapping,

$$I_{out} = I_{out,\min} + (I_{out,\max} - I_{out,\min})\frac{(\beta_{in} - \beta_{in,\min})}{(\beta_{in,\max} - \beta_{in,\min})}, \qquad (1)$$

with fixed mapping parameters equal to $\beta_{in,\min} = 5.0 \times 10^{-11}$, $\beta_{in,\max} = 7.0 \times 10^{-10}$, $I_{out,\min} = 0$ and $I_{out,\max} = 2^{12} - 1 = 4095$, with the last two parameters representing the lowest and highest pixel intensity values in the output rescaled 12-bit images. All input values smaller than $\beta_{in,\min}$ were converted to $\beta_{in,\min}$, and all input values larger than $\beta_{in,\max}$ were converted to $\beta_{in,\max}$, prior to the linear mapping. The values $\beta_{in,\min} = 5.0 \times 10^{-11}$ and $\beta_{in,\max} = 7.0 \times 10^{-10}$ were chosen to be just below the lowest possible value of $\beta$ for soft breast tissues (at *E* = 32 keV) and above the highest possible value of $\beta$ for calcifications, respectively. Therefore, any pixel value outside the range $[\beta_{in,\min}, \beta_{in,\max}]$ could appear in the reconstructed floating-point images only due to noise, artefacts or extraneous inclusions, such as e.g. surgical clips. Choosing this input range for the linear mapping allowed us to effectively maximise the contrast of the essential features in the output 12-bit images, without losing any important information.

### e. Objective assessment

Objective assessment of CT-reconstructed coronal and axial slices was performed using the contrast, spatial resolution and signal-to-noise ratio (SNR) as measures of image quality. It was shown in (Gureyev et al., 2014; Gureyev et al., 2016) that, in the case of CT imaging at a fixed radiation dose, the ratio of SNR$^2$ to spatial resolution in the third power is proportional to Shannon's information capacity of the imaging system. In other words, this ratio reflects the capacity of the system to extract information from each detected photon about the 3D distribution of the refractive index on which the photons were scattered. It is also known that, when an image is post-processed by means of linear filtering (e.g. convolution or deconvolution), its SNR and spatial resolution change. For example, low-pass filtering typically increases the SNR, but spoils the spatial resolution. However, the ratio of SNR to spatial resolution (in the appropriate power, depending on the dimensionality of the image) remains constant after such image filtering operations (Gureyev et al., 2016) and can be changed only by means of non-linear processing such as, for example, Machine Learning or other methods utilising a priori information about the imaging system or the imaged object.

The contrast was measured in the images by selecting rectangular regions across boundaries between the adjacent adipose and glandular tissue areas, 1D-averaging the pixel values along the shorter dimension of the rectangle and creating a histogram from the resultant averaged 1D profile along the longer dimension of the rectangle. The histogram was created by dividing the range of all 1D-averaged pixel values into five equal intervals and creating five bins containing the pixels with values in the corresponding intervals. The contrast was then defined as

$$C = \frac{\bar{\beta}_{\max} - \bar{\beta}_{\min}}{\bar{\beta}_{\max} + \bar{\beta}_{\min}}, \quad (2)$$

where $\bar{\beta}_{\max}$ and $\bar{\beta}_{\min}$ were the average values of the top and bottom bins, respectively, of the 5-bin histogram. Note that in eq.(2) and in subsequent related formulae below, we used the notation corresponding to the original reconstructed coronal slices which contained 2D distributions of 32-bit floating-point values of the imaginary part of the refractive index, $\beta$. When similar measurements were performed on the 3mm thick axial slices, then, instead of $\beta$, the relevant formulae involved the corresponding 12-bit pixel intensity values, $I$, obtained according to eq.(1).

The SNR of a stochastic function, such as e.g. reconstructed distribution of refractive index, $\beta(\mathbf{r})$ at a point **r**, was defined as

$$\mathrm{SNR} = \frac{\bar{\beta}(\mathbf{r})}{\sigma_\beta(\mathbf{r})}, \quad (3)$$

where $\bar{\beta}(\mathbf{r})$ was the statistical mean ("the signal") and $\sigma_\beta^2(\mathbf{r}) \equiv \overline{[\beta(\mathbf{r}) - \bar{\beta}(\mathbf{r})]^2}$ was the (noise) variance. In the present study, practical measurements of SNR were performed in "flat" regions of reconstructed slices. In this context, the flatness of an image inside a certain region means that the reconstructed values of $\beta(\mathbf{r})$ were approximately constant in that region. We generally assumed that the "spatial ergodicity" hypothesis was satisfied in our images, implying that the ensemble averages, involved in eq.(2) and elsewhere, could in practice be substituted by spatial averages within such flat areas (Goodman, 2000). This allowed us to evaluate the SNR from single stochastic images instead of ensembles of such images. Evaluating the SNR in flat areas of the reconstructed slices also allowed us to minimise the contribution of the natural variations of the tissue density to the measured values of SNR.

The contrast-to-noise (CNR), was defined according to the following formula:

$$\text{CNR} = C \times \text{SNR} = \frac{\bar{\beta}_{\max} - \bar{\beta}_{\min}}{2\sigma_\beta(\mathbf{r})} \frac{\bar{\beta}(\mathbf{r})}{(\bar{\beta}_{\max} + \bar{\beta}_{\min})/2} \qquad (4)$$

Note that the CNR defined in this way was dimensionless and invariant with respect to linear scaling of $\beta$. Furthermore, while the contrast $C$ in eq.(4) was measured as described in conjunction with eq.(2) above, the SNR was measured outside the selected rectangle crossing the boundary between adjacent glandular and adipose tissue regions (in which the contrast was evaluated). This was done in order to avoid the influence of the strong tissue density variation inside the selected rectangle on SNR. The square region where the SNR was evaluated (according to eq.(3)) in conjunction with eq.(4) was always selected immediately adjacent to the top or the left side of the rectangle, while ensuring that the region where SNR was measured lied in a flat area of the image.

Since we are interested in estimating the spatial resolution in imaging, we focus on the spatial resolution of the corresponding class of 3D computational imaging systems that is relevant to the present study. Such imaging systems incorporate illumination of a sample with coherent monochromatic incident X-ray beam, transmission (scattering) of the beam through the sample, free-space propagation of the transmitted beam from the sample to the detector, image acquisition by the detector and, finally, a CT reconstruction of the 3D distribution of refractive index in the sample. The spatial resolution of such composite hardware-software imaging systems is defined as the width of the 3D point-spread function (PSF) of the imaging system, i.e. the width of response of the imaging system to a delta-function-like input (with an idealised infinitely-narrow 3D point-like sample). Assuming that the imaging system is linear and shift-invariant, any real (reconstructed) image is considered to be a convolution of an "ideal reconstruction", $\beta_{id}(\mathbf{r})$, corresponding to an imaging system with a delta-function PSF, and a real PSF, $P(\mathbf{r})$, of the imaging system: $\beta(\mathbf{r}) = (\beta_{id} * P)(\mathbf{r}) \equiv \int \beta_{id}(\mathbf{r}')P(\mathbf{r}-\mathbf{r}')d\mathbf{r}'$. In turn, the width of the 3D PSF is defined via its second spatial integral moment:

$$\text{Res} \equiv \left( \frac{4}{3} \frac{\int |\mathbf{r} - \bar{\mathbf{r}}|^2 P(\mathbf{r})d\mathbf{r}}{\int P(\mathbf{r})d\mathbf{r}} \right)^{1/2}, \qquad (5)$$

where $\bar{\mathbf{r}} \equiv \int \mathbf{r} P(\mathbf{r})d\mathbf{r}$. Regarding the choice of normalization factor, (4/3), included in eq.(5), note that, in the case of n-dimensional Gaussian distributions, $P_{Gauss}(\mathbf{r}) = (2\pi)^{-n/2}\sigma^{-n}\exp[-|\mathbf{r}|^2/(2\sigma^2)]$, we get $\text{Res} = 2\sigma$.

As mentioned above, an objective imaging quality characteristic (Gureyev et al., 2016), closely related to Shannon's information capacity of the imaging system, is proportional to the ratio of SNR to the appropriate power of the spatial resolution. For a 3D imaging system, such as CT, the relevant ratio is $\text{SNR}^2/\text{Res}^3$ (Gureyev et al., 2016). Note that in the case of Poisson photon-counting statistics, the latter ratio corresponds to the number of photons per minimal resolvable volume. It is therefore clear, in particular, that $\text{SNR}^2/\text{Res}^3$ is proportional to the incident photon fluence and, hence, is also proportional to the radiation dose, $D$, delivered to the sample during imaging. Accordingly, the ratio

$$Q_S \equiv \frac{\text{SNR}}{\text{Res}^{3/2}D^{1/2}}, \qquad (6)$$

which is closely related to "intrinsic imaging quality characteristic" (Gureyev et al., 2014; Gureyev et al., 2016), reflects the amount of Shannon information that the imaging system is able to extract per one incident photon. As in the present study we compared the quality of three different PB-CT reconstruction algorithms using the data from scans collected at a fixed dose (2 mGy MGD), we ignored the constant dose parameter in our objective comparisons and only measured and reported the ratios of $\text{SNR}/\text{Res}^{3/2}$. Measurements of spatial resolution were also performed in "flat" regions of reconstructed slices. Evaluation of the spatial resolution was based on the effect of PSF on the noise distribution in images (Goodman, 2000). We used a method based on the Fourier transform of the equation $\beta(\mathbf{r}) = (\beta_{id} * P)(\mathbf{r})$:

$$\hat{\beta}(\mathbf{k}) = \hat{\beta}_{id}(\mathbf{k})\hat{P}(\mathbf{k}), \qquad (7)$$

where the overhead hat symbol denotes the Fourier transform, $\hat{f}(\mathbf{k}) = \iint \exp(-2\pi \mathbf{k} \cdot \mathbf{r}) f(\mathbf{r}) d\mathbf{r}$. The noisy photon fluence in the ideal reconstruction was assumed to be uncorrelated between different voxels and having constant mean and variance within flat regions. Then the Fourier transform of the ideal reconstruction was also a flat noisy distribution and the width of the product of the two functions in the right-hand side of eq.(7) was determined primarily by the width of the modulation transfer function (MTF), $|\hat{P}(\mathbf{k})|$. The width of the MTF was straightforward to measure in practice using Fourier transforms of flat regions of reconstructed coronal slices of $\beta(\mathbf{r})$ in accordance with eq.(5) with $|\hat{P}(\mathbf{k})|$ in place of $P(\mathbf{r})$. After that, assuming that the PSF was approximately Gaussian, and hence the MTF was also Gaussian, we applied the known relationship between the widths of a Gaussian distribution and its Fourier transform to evaluate the width of the PSF (Gureyev et al., 2023):

$$\text{Res}[P_{Gauss}] = 2/\text{Res}[\hat{P}_{Gauss}]. \qquad (8)$$

For each objective metric, statistical tests were conducted to evaluate significant differences among the reconstruction algorithms. A paired sample t-test was performed using MATLAB 2020a with the Statistical and Machine Learning Toolbox. To account for multiple comparisons among the three algorithms, the Bonferroni correction was applied. This method adjusts the significance level to control for the increased risk of "False positive" errors when performing multiple tests. Since three pairwise comparisons were conducted (cSART vs. UTR, cSART vs. FFBP, UTR vs. FBP), the significance level was divided by the number of comparisons ($\alpha$ = 0.05/3), resulting in an adjusted $\alpha$ = 0.017.

    f. Subjective assessment

This assessment was performed on 3 mm-thick craniocaudal (axial) slices reconstructed using the three proposed algorithms. Seven assessors independently evaluated the radiological image quality in this study. The panel of assessors comprised breast specialist radiologists, diagnostic radiographers and medical physicists. Conducted in a setup resembling digital mammography reading rooms, the assessments employed a high-specification workstation equipped with a single 12MP monitor and typical tools such as zooming, panning, and window/level adjustment. The evaluation process involved examining the image quality of three different reconstructed techniques displayed in a three-panel (grids) in a synchronised format. Here, the middle panel served as the reference against which the right and left panels were compared. To minimize order effects and sequence bias, the image sets within the panels were randomly allocated for each sample, with assessors unaware of the sequence of the image sets (reconstruction techniques) in each panel.

Using a five-point rating scale, assessors were tasked with evaluating image quality criteria for each sample. This scale ranged from indicating clearly better (+2) or slightly better (+1) image quality compared to the reference, to equal quality (0), slightly worse (-1), or clearly worse (-2) than the reference. For each sample, assessors were asked to rate the following five Image Quality Criteria.

- Perceptible Contrast: the differences in radiolucency between soft tissue regions, indicating how well soft tissue variations were displayed.
- Sharpness of Tissue Interfaces: the clarity of boundaries between different tissue types, measuring how well the image visualized transitions between tissues.
- Calcification Visibility: the visibility and sharpness of micro-calcifications, specifically assessing their clarity and prominence in the image.
- Image Noise: the presence of quantum mottle in the image, with a higher score indicating less noise in the test image compared to the reference.
- Overall Image Quality: a holistic assessment of the entire stack of images, considering all aspects of image quality.

The observer study employed a Multiple-Reader, Multiple-Case (MRMC) design, where all assessors evaluated all images. Inter-observer agreement regarding image ratings was assessed using the intraclass correlation coefficient (ICC). Utilizing SPSS Statistics v28, a two-way mixed model of ICC was generated based on absolute rating scores. It's widely acknowledged that an ICC below 0.4 suggests poor reliability, while values between 0.4 and 0.6 indicate fair reliability, 0.6 to 0.75 imply good reliability, and anything exceeding 0.75 reflects excellent reliability. We also examined the agreement between assessors by calculating Cronbach's Alpha for all assessors and performed sensitivity analyses by individually excluding each assessor to ensure there were no outliers in the panel of assessors.

Following this, image quality underwent analysis through visual grading characteristics (VGC) analysis using VGC Analyzer software v1.0.2. For each image criterion, the cumulative distributions of rating data for the test images were plotted against the reference images, yielding a curve. The area under this curve (AUCVGC) served as a metric for measuring the difference in image quality between the two sets. In the interpretation of VGC analysis results, an AUCVGC of 0.5 denotes equivalence between the image sets, values between 0 and 0.5 signify lower quality and those between 0.5 and 1 indicate higher quality in the test images compared to the reference.

Statistical tests employed a nonparametric approach. To establish the confidence interval (CI) of the AUCVGC and calculate p-values for testing the null hypothesis, bootstrapping was conducted with 2000 resamplings of the rating scores. Given the small number of assessors, the analysis considered a random-observer scenario, incorporating bootstrapping of assessors to ensure results' generalizability to the assessor population.

### 3. Results and Discussion
#### a. Objective image quality measurements

Figure 1 depicts three coronal slices from sample 5, reconstructed using the three considered algorithms. Signal-to-noise ratio (SNR) and spatial resolution were measured within uniform regions selected in adipose tissue (see Fig.1a). Contrast was measured across fibroglandular and adipose interfaces (see Fig.1b). The objective image quality measurements were performed on two different datasets: the original reconstructions in the form of coronal slices and the axial slices obtained with the 30-pixel binning described in the previous section. For the reconstructions of the coronal slices, which retained the original voxel size, regions of interest (ROIs) of 128 × 128 pixels were used for the measurements of SNR and spatial resolution. For the thick axial slices, we employed ROIs of 64 × 64 pixels, as the selection of uniform regions was more challenging after the binning procedure which increased the apparent presence of glandular tissue. All the measurements were made using the definitions provided in eqs. (2)-(8), as implemented in the X-TRACT software (Gureyev et al., 2011). For each sample, three slices were selected to evaluate the objective metrics for each reconstruction algorithm. The primary criterion for selecting the region of interest was the presence of a sufficiently large and uniform area to accommodate the ROIs of the chosen size within adipose tissue. Results of the analysis of thin coronal slices are presented in Figure 2.

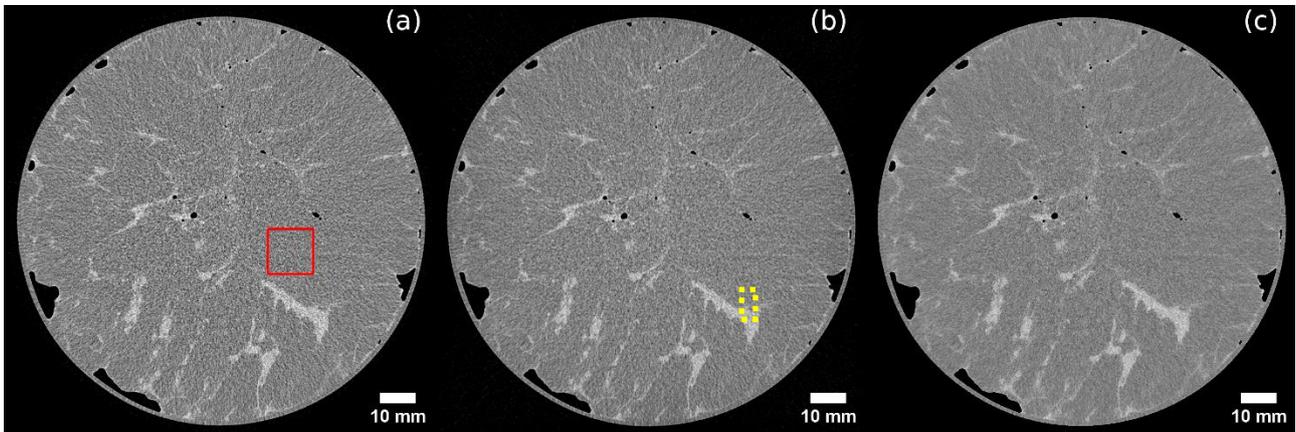

*Figure 1. Coronal slices from sample 5, reconstructed using the three considered algorithms: (a) the image obtained with FBP, (b) UTR, (c) cSART. The solid red square highlights a region of interest used for the measurement of SNR and spatial resolution, while the yellow dotted rectangle shows a typical selection used for measurement of contrast and CNR.*

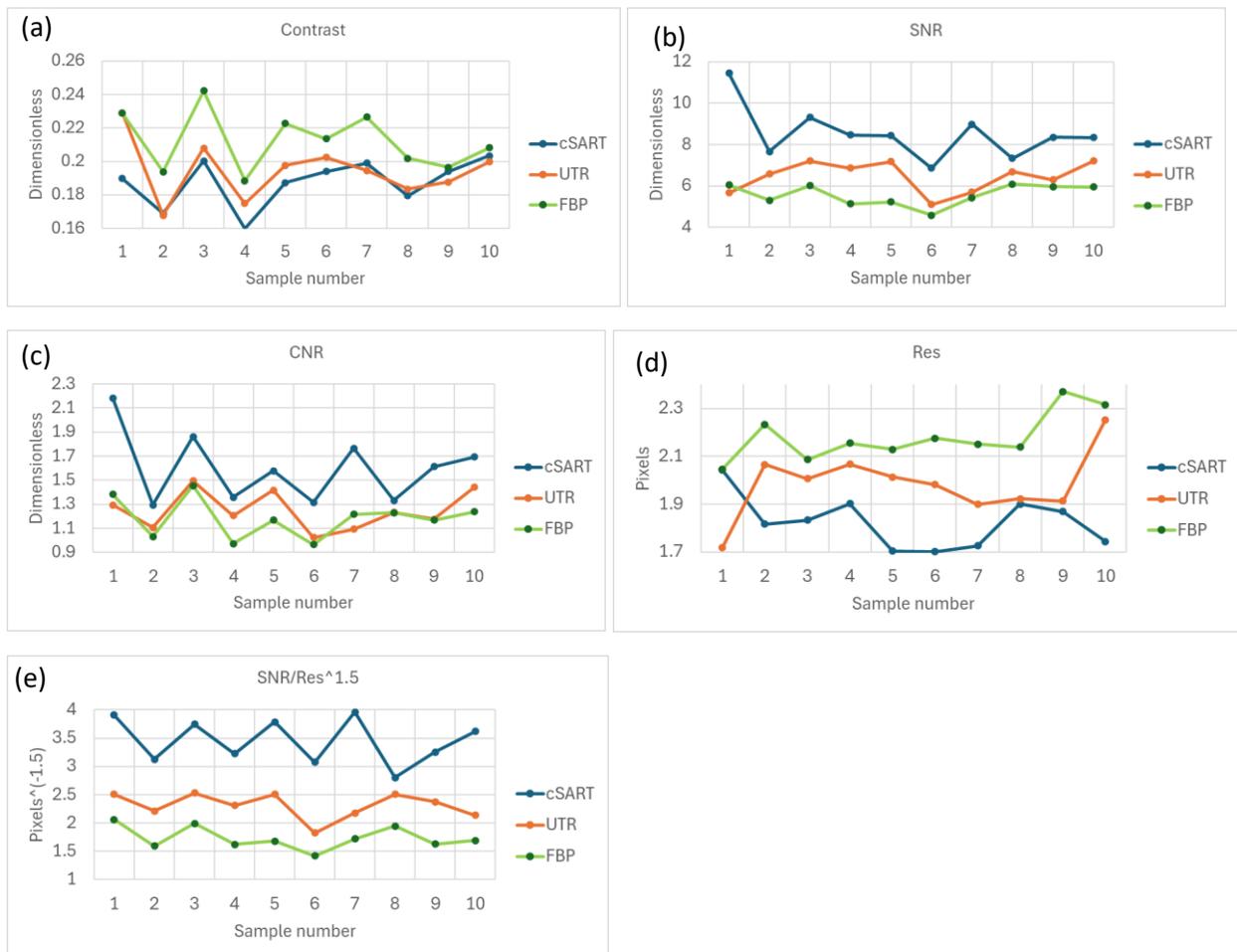

*Figure 2. Objective measurements of image quality characteristics in coronal slices of 10 different mastectomy samples reconstructed using the cSART, UTR and FBP algorithms from PB-CT scans collected with 32 keV planar monochromatic X-rays and the MGD of 2 mGy. (a) Contrast, (b) SNR, (c) CNR, (d) Spatial resolution (in the detector pixel size), (e) SNR/Res$^{1.5}$ (in the inverse detector pixel size taken to the power 1.5).*

Figure 2(a) shows the contrast values, measured according to Eq. (2), in thin coronal slices of 10 samples reconstructed using the three algorithms. The contrast was generally highest in the FBP-reconstructed images (p-values against UTR and cSART < 0.001), while UTR and cSART produced lower and statistically indistinguishable contrast (p-value = 0.14).

Figure 2(b) presents the SNR values, measured in the same slices according to Eq. (3). The results demonstrate that cSART consistently achieved the highest SNR values (minimum = 6.9, maximum = 11.5, average = 8.5), followed by UTR (minimum = 5.1, maximum = 7.2, average = 6.4), with FBP performing the worst (minimum = 4.6, maximum = 6.1, average = 5.6). These differences were statistically significant, as evaluated by paired samples t-tests (all p-values < 0.01). Notably, the minimum, maximum, and average values reported were calculated across all samples, while the individual data points in the plots represent the average values from the three slices per sample.

Figure 2(c) depicts the CNR values, measured according to Eq. (4), for the same 10 samples. Similar to the SNR results, cSART yielded the highest CNR values, while UTR and FBP produced comparable results. Statistical analysis confirmed significant differences between cSART and the other algorithms (p-values < 0.01), but no significant difference between UTR and FBP (p-value = 0.13).

The spatial resolution, measured using Eqs. (5), (7), and (8), is shown in Figure 2(d). cSART provided the best spatial resolution (minimum = 170 µm, maximum = 204 µm, average = 182 µm), while FBP exhibited the worst performance (minimum = 205 µm, maximum = 237 µm, average = 218 µm). UTR results fell in between (minimum = 172 µm, maximum = 225 µm, average = 198 µm). Paired samples t-tests produced the following p-values: cSART vs. UTR = 0.05 (not significant), cSART vs. FBP and UTR vs. FBP < 0.001.

Finally, Figure 2(e) presents the calculated proxy for intrinsic imaging quality, as defined by Eq. (6), which involves the ratio of SNR to the resolution raised to the power of 1.5. This metric clearly favoured cSART, with UTR as the second best and FBP performing the worst (all p-values < 0.001). These results indicate that cSART-reconstructed slices objectively contained more measurable (Shannon) information about the imaged samples compared to slices reconstructed using UTR or FBP.

Figure 3 presents an example of three 3 mm-thick craniocaudal (axial) slices from sample 6, reconstructed using the three proposed algorithms. A large lesion with multiple micro-calcification clusters is clearly visible in the upper-central region of these slices.

Figure 4 presents the measurements of objective image quality characteristics in 3 mm-thick, 12-bit axial slices. Each data point represents the average value calculated across three selected slices. Overall, the results observed in these thick axial slices closely mirror those obtained for the thin coronal slices presented earlier. Repeating the measurements in the rescaled thick axial slices was crucial, as these slices (rather than the thin coronal ones) were used for the subjective image quality assessments described in the subsequent section.

Figure 4(a) shows that the contrast was significantly highest in the FBP-reconstructed slices (p-values < 0.001). While cSART and UTR produced generally comparable contrast values, the UTR contrast was statistically higher than that of cSART (p-value < 0.01).

Figure 4(b) demonstrates that the SNR results align with those found for the coronal slices. Specifically, cSART consistently achieved the highest SNR values (minimum = 8.7, maximum = 12.8, average = 10.9), followed by UTR (minimum = 6.3, maximum = 10.6, average = 8.3), and FBP yielding the lowest values (minimum = 4.8, maximum = 8.6, average = 6.3). These differences were highly statistically significant, as determined by paired samples t-tests (all p-values < 0.001).

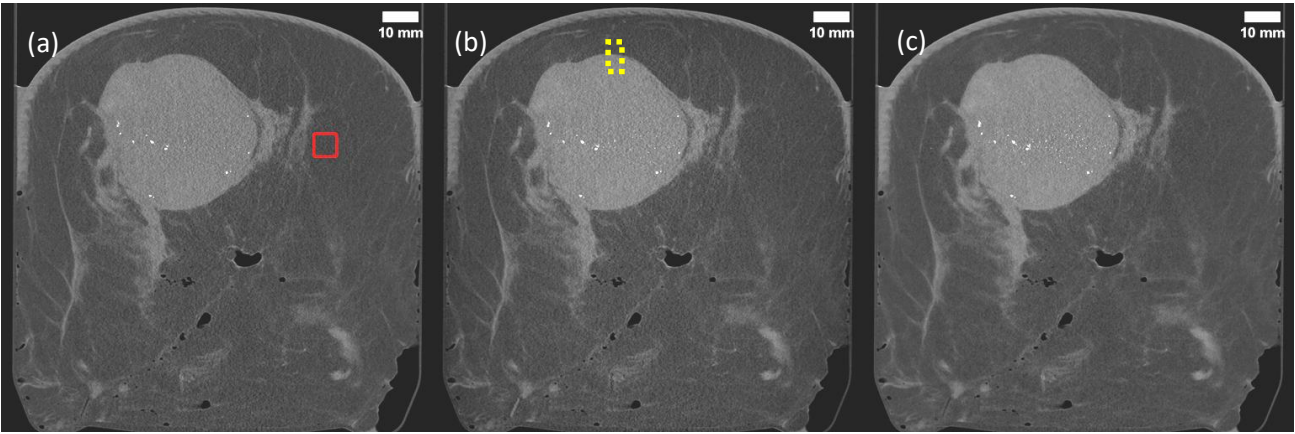

*Figure 3. Reconstructed 3mm-thick axial slices from sample 6, obtained with the three considered algorithms: (a) the image obtained with FBP, (b) UTR, (c) cSART. The solid red square highlights a region of interest used for the measurement of SNR and spatial resolution, while the yellow dotted rectangle shows a typical selection used for measurement of contrast and CNR.*

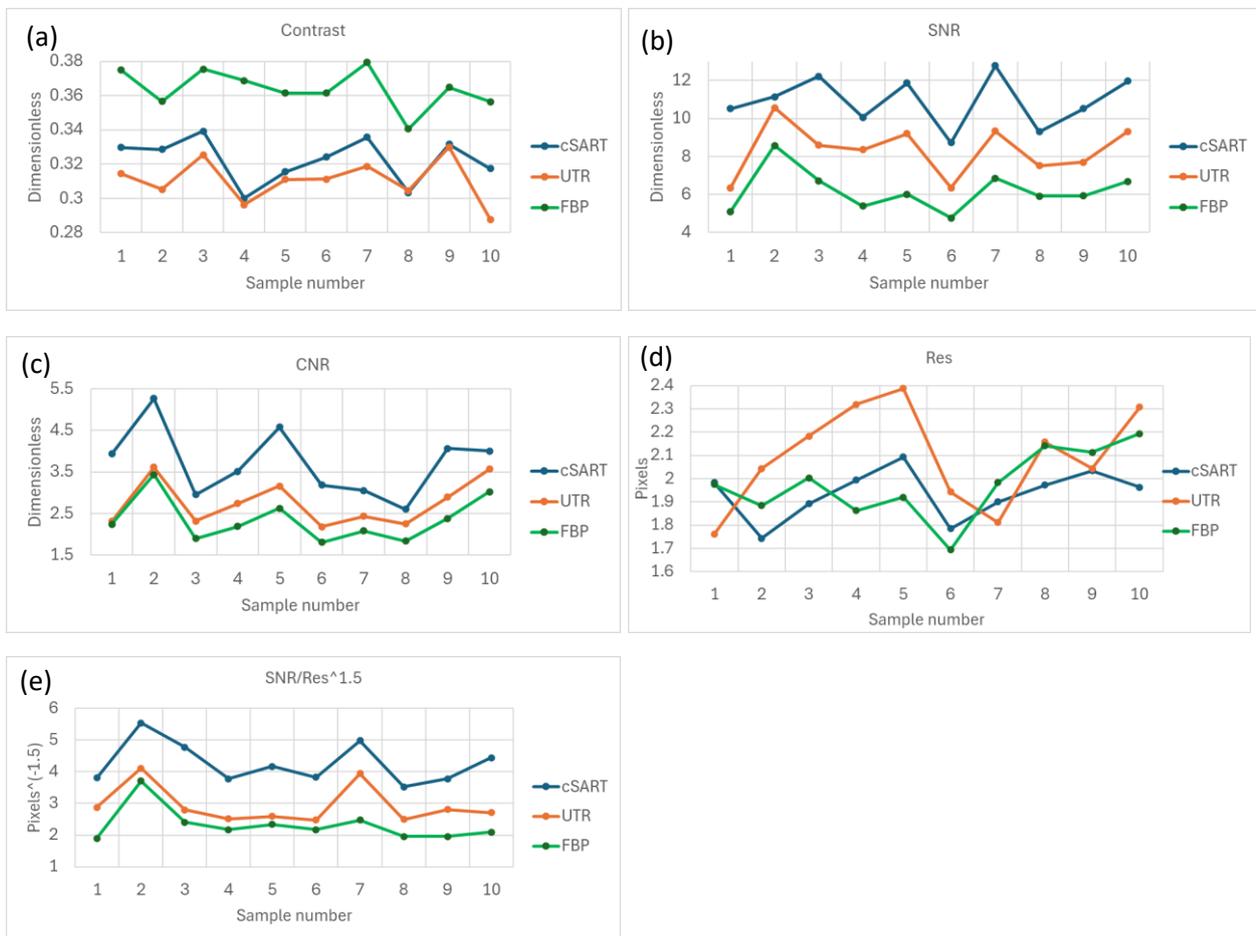

*Figure 4. Objective measurements of image quality characteristics in 3mm-thick axial slices of 10 different mastectomy samples reconstructed using the cSART, UTR and FBP algorithms from PB-CT scans collected with 32 keV planar monochromatic X-rays and the MGD of 2 mGy. (a) Contrast, (b) SNR, (c) CNR, (d) Spatial resolution (in the detector pixel size), (e) SNR/Res$^{1.5}$ (in the inverse detector pixel size taken to the power 1.5).*

The trends in CNR, shown in Figure 4(c), mirrored those observed for SNR. cSART exhibited the highest CNR values, followed by UTR and FBP, with all comparisons demonstrating statistical significance (p-values < 0.001).

In contrast, the spatial resolution results measured in the thick axial slices (Figure 4(d)) revealed no statistically significant differences between the three algorithms. The p-values were 0.03, 0.36, and 0.15 for cSART vs. UTR, cSART vs. FBP, and UTR vs. FBP, respectively. However, when the proxy for intrinsic imaging quality - defined as the ratio of SNR to spatial resolution raised to the power of 1.5 - was analyzed (Figure 4(e)), the results clearly differentiated the three algorithms. This differentiation was driven primarily by the significant statistical differences observed in SNR (all p-values < 0.001).

### b. Subjective assessment of image quality

The results of the subjective image quality assessment are summarized in Figure 5 and Table 2. The calculated ICC for this study indicated excellent agreement among assessors for rating perceptible contrast, sharpness of tissue interfaces, and image noise, with good reliability observed in the ratings of calcification visibility and overall image quality. Note that some features in the images that looked like calcifications were in fact the artefacts from brighter noisy pixels that have been amplified by our thresholding algorithm (see e.g. Fig.3(c)). Despite the appearance of these artefacts in a small number of images, we still consider our current thresholding algorithm (used for conversion from the 0.1 mm thick 32-bit floating-point coronal slices to 3-mm thick 12-bit integer axial slices) the best compromise preventing the calcifications from being "washed out" by the 30-pixel averaging involved in the conversion process. While the presence of these artefacts may have affected the evaluation of calcification visibility in the subjective assessment, the effect is not expected to be a major one, as only a few images were affected by this problem.

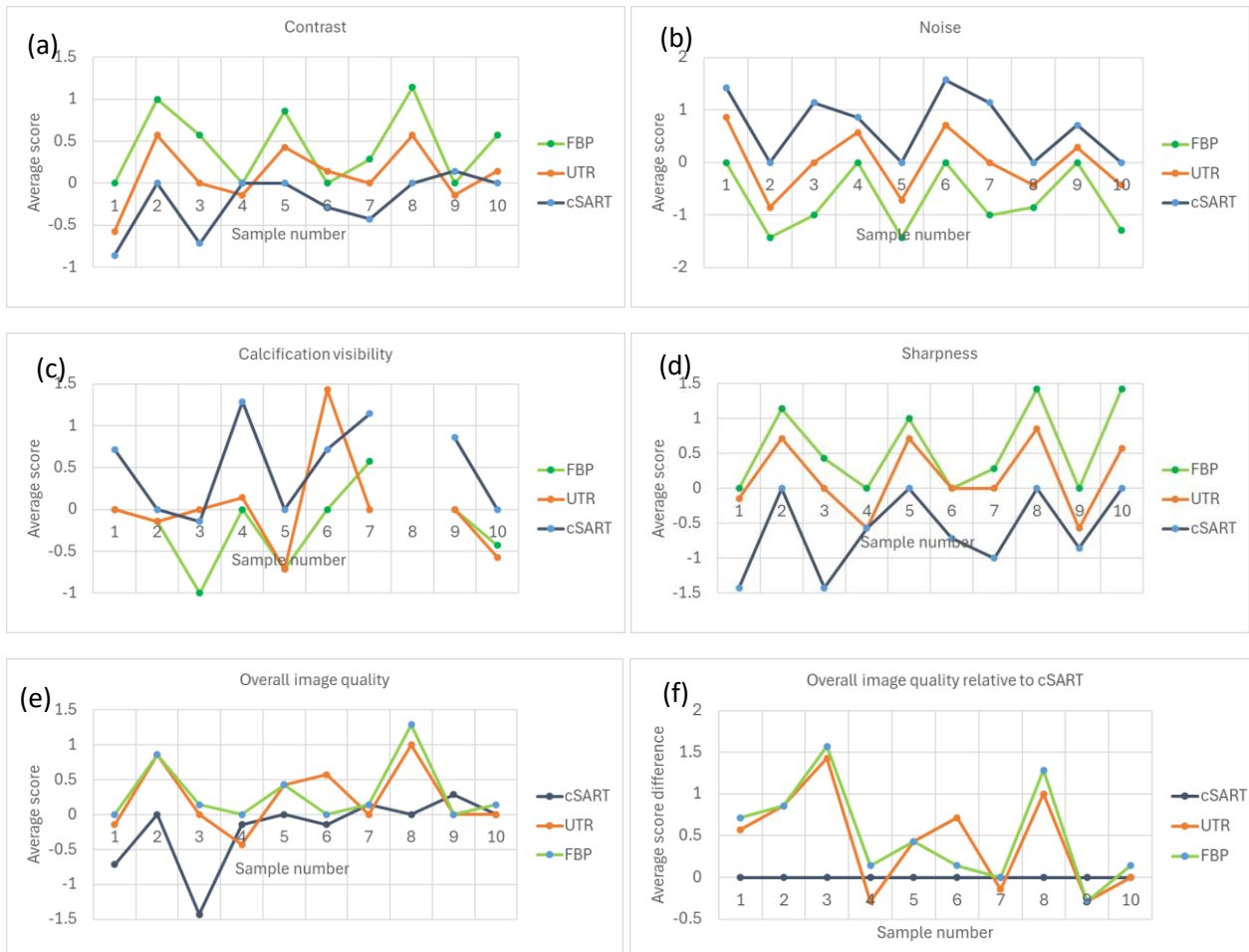

*Figure 5. Subjective assessment of image quality in 3mm-thick axial slices of 10 different mastectomy samples reconstructed using the cSART, UTR and FBP algorithms from PB-CT scans collected with 32 keV planar monochromatic X-rays and the MGD of 2 mGy. (a) Average (across all assessors) subjective image contrast scores; (b) average image noise scores; (c) average calcification visibility scores (note that sample 8 did not contain any calcifications); (d) average image sharpness scores; (e) average overall image quality scores; (b) differences between the average overall image quality scores for each of the three algorithm and those of cSART.*

When assessing contrast and sharpness, FBP received the highest scores from the assessors, followed by UTR. FBP demonstrated significantly better performance than both UTR (p = 0.01) and cSART (p = 0.01) for both criteria, while the difference between UTR and cSART was significant for sharpness (p < 0.01) but not for contrast. In terms of calcification visibility, cSART achieved the highest scores, followed by UTR; however, the only significant difference among the three techniques was observed between cSART and FBP. For image noise, cSART was rated significantly better than both FBP (p < 0.01) and UTR (p < 0.01), while UTR was significantly better than FBP (p < 0.01). Regarding overall image quality, which encompasses various image quality criteria, there was no significant difference between FBP and UTR, but both of these techniques were rated significantly better than cSART (p = 0.02 and p = 0.03, respectively).

*Table 2. Results of assessment of subjective image quality in 3mm-thick axial slices of 10 different mastectomy samples reconstructed using the cSART, UTR and FBP algorithms from PB-CT scans at E = 32 keV X-rays at 2 mGy MGD.*

| Image Quality Criteria | Inter-observer ICC | Average image quality score | | | VGCAUC and p-value | | |
|---|---|---|---|---|---|---|---|
| | | FBP | UTR | cSART | UTR against FBP | cSART against FBP | cSART against UTR |
| Perceptible Contrast | 0.80 | 0.4 | 0.1 | -0.2 | **0.38 (0.01)** | **0.29 (0.01)** | 0.39 (0.06) |
| Sharpness of Tissue Interfaces | 0.90 | 0.6 | 0.2 | -0.6 | **0.38 (0.01)** | **0.17 (<0.001)** | **0.26 (<0.01)** |
| Calcification Visibility | 0.68 | -0.2 | 0 | 0.5 | 0.55 (0.35) | **0.68 (0.02)** | 0.62 (0.11) |
| Image Noise | 0.91 | -0.7 | 0 | 0.7 | **0.70 (<0.01)** | **0.82 (<0.01)** | **0.70 (<0.01)** |
| Overall Image Quality | 0.62 | 0.3 | 0.2 | -0.2 | 0.46 (0.31) | **0.32 (0.02)** | **0.35 (0.03)** |

c. Discussion

While the results of the objective image quality measurements clearly show that cSART consistently outperforms both UTR and FBP in terms of various quality characteristics, including spatial resolution, SNR, and CNR, the subjective image quality assessment revealed a general preference for FBP reconstructions. Interestingly, the only objective metric that strongly correlated with the subjective evaluation was the image contrast, which was consistently higher for FBP compared to both UTR and cSART. This outcome suggests that radiologists may place greater importance on image contrast (i.e., the sharpness and visibility of key image features) when subjectively assessing image quality, compared to other factors such as image noise (SNR), which is a major component of objective assessments.

This observation leads to a hypothesis, which could be explored further in future studies: medical imaging specialists, when evaluating medical images, may subjectively assign more weight to contrast, particularly in terms of visual clarity and feature recognition, rather than noise reduction, even though the latter is objectively critical for image quality. The results of the current study indicate that cSART, in terms of objective metrics, is more capable of providing Shannon information about the imaged samples., This suggests that cSART could be a better choice for automated image analysis tools, especially with the rapid growth of AI-assisted diagnostics. However, if subjective preferences for contrast persist in clinical practice, this may influence algorithm selection in certain settings, such as visual assessments by radiologists.

Despite this difference between objective and subjective assessments, we argue that the findings of this study do not diminish the value of the cSART and UTR reconstruction algorithms. Rather, they highlight the complexity of image quality evaluation, where both objective metrics and subjective radiological preferences play significant roles. If automated tools that rely on objective quality metrics gain further acceptance in medical imaging, cSART's superior information content could become a major advantage for applications in breast phase-contrast CT imaging.

Furthermore, beyond these results, the current work presents a robust methodology for medical image quality assessment that extends beyond the specific context of phase-contrast breast CT imaging. We proposed a systematic use of a set of five physics-based objective image quality characteristics: contrast, SNR, CNR, spatial resolution, and Qs, with the latter characteristic particularly effective in quantifying the Shannon information about the imaged sample provided by a computational imaging system. Complementing these objective metrics, we also applied our previously developed methodology for systematic subjective image quality assessment, using five key characteristics—Perceptible Contrast, Sharpness of Tissue Interfaces, Calcification Visibility, Image Noise, and Overall Image Quality—and analyzed them through Visual Grading Characteristics (VGC). This integrated approach, which has been developed in collaboration with practicing radiologists and medical imaging specialists, provides a comprehensive way to assess medical image quality and to explore the relationships between objective and subjective image quality evaluations.

By combining both objective and subjective assessments, this study allows for an in-depth analysis of the strengths and limitations of each of the three PCT reconstruction approaches and highlights the potential for these findings to inform the development of automated image analysis tools in the future.

## Acknowledgements

The experimental part of this research was undertaken on the Imaging and Medical beamline at the Australian Synchrotron, part of ANSTO. The following funding is acknowledged: National Health and Medical Research Council, Australia (APP2011204). The authors would like to acknowledge staff from the University of Melbourne, Monash University, Australian Synchrotron and the University of Sydney for their contribution toward the collection of the original image data and the radiological assessment. Sandro Donato is supported by the project Tech4You- Technologies for climate change adaptation and quality of life improvement (C.I. ECS 00000009, CUP H23C22000370006)